\documentclass[doublecol]{epl2} 
\usepackage{amsmath,amssymb}

\newcommand \be {\begin{equation}}
\newcommand \ee {\end{equation}}
\newcommand \bea {\begin{eqnarray}}
\newcommand \eea {\end{eqnarray}}

\title{Socio-economic utility and chemical potential}

\author{R\'emi Lemoy\inst{1,2} \and Eric Bertin\inst{2,3} \and Pablo Jensen\inst{1,2,3}}
\shortauthor{R. Lemoy \etal}

\institute{                    
  \inst{1} LET Transport Economics Laboratory - Universit\'e Lyon-2 and CNRS, 14 Avenue Berthelot, F-69007 Lyon, France\\
  \inst{2} IXXI Complex Systems Institute - 5 rue du Vercors F-69007 Lyon, France\\
  \inst{3} Universit\'e de Lyon, Laboratoire de Physique, ENS Lyon, CNRS, 46 All\'ee d'Italie, F-69007 Lyon, France}


\abstract{In statistical physics, the conservation of particle number results in the
equalization of the chemical potential throughout a system at equilibrium.
In contrast, the homogeneity of utility in socio-economic models is usually thought to rely on the competition between individuals, leading to Nash equilibrium.
We show that both views can be reconciled by introducing a notion of chemical potential
in a wide class of socio-economic models, and by relating it in a direct way to
the equilibrium value of the utility. This approach also allows the dependence
of utility across the system to be determined when agents take decisions
in a probabilistic way. Numerical simulations of a urban economic model also suggest that our result is valid beyond the initially considered class of solvable models.
}

\begin{document}

\maketitle

Socio-economic sciences and statistical physics are both interested in the evolution of systems characterized by a large number of interacting entities. These entities can for instance be economic or social agents in social sciences \cite{smith,schelling71,latour07}, atoms or molecules in statistical physics \cite{cambridge,goodstein,balescu}. The question of the emergence of macroscopic patterns from the interactions of a large number of microscopic agents is studied by both fields of science.
In statistical physics, a quantitative framework has been developed over the last century, allowing the equilibrium behaviour of large assemblies of atoms or molecules to be handled precisely \cite{balescu}. 

In socio-economic models, the preferences of individuals are usually characterized by a utility function, which describes their welfare with respect to their current situation or environment.
Each individual or agent wants to maximize his own welfare. Decisions (e.g., moving to a more convenient place) are thus taken in a purely selfish way, while in physics the motion of particles
is governed by the variation of the total energy.
Recently, a global function linking individual decisions to the variation of a global quantity has been introduced to describe some classes of socio-economic models \cite{pnas2009,jensen2009}. This approach then allows such models to be described with statistical physics tools. Importantly, the equilibrium state can then be calculated by maximizing a state function (akin to a free energy) instead of having to solve a complicated Nash equilibrium of strategically interacting agents.

The question we investigate in this letter is whether this physical description of socio-economic models can be extended to other basic concepts of statistical physics, such as the equalization of thermodynamic parameters like temperature or chemical potential.
The equalization of these quantities throughout the system precisely results
from the conservation of the conjugated extensive quantities, namely the energy
or the number of particles.
Although there is no notion of energy in socio-economic models, the dynamics indeed conserves
the number of agents. A natural question is thus to know whether a chemical potential can be
defined in such models, and what would be its relation to
standard socio-economic concepts.
This question is further motivated by the following remark.
In spatial socio-economic models, the individual dynamics leads to a Nash equilibrium,
where no agent has an incentive to move. If all agents are of the same type,
the Nash equilibrium results in a spatially uniform utility, even if the environment
is spatially inhomogeneous like in cities, where the center plays a specific role.
This uniformity property is also expected from the chemical potential (if such a quantity
can be defined), suggesting a possible relation between these two notions.

Here, we investigate this issue in the framework of a generic class of exactly
solvable models involving a population of locally interacting agents.
We define in a precise way a chemical potential for this class of models,
and provide a direct link between the chemical potential
and the socio-economic utility.
Two explicit examples from the field of urban economics are also presented.

%%%%%%%%%%%%%%%%%%%%%%%%%%%%%%%%%%%%%%%%%%%%%%%%%%%

\section{Model and dynamics}
This work deals with socio-economic models characterized by a large number of interacting agents,
residing on a set of sites, labeled by an index $q=1,\ldots,Q$.
Agents are able to move from one site to another in order to increase their utility.
In addition, agents belong to $m$ different groups, according for instance
to their income, or to their cultural preferences.
The variables used to describe the system are the numbers $n_{qi}$ of agents
of each group $i=1,\ldots,m$ at each node $q$. The configuration of the
system is described by the set $x=\{n_{qi}\}$.
We assume that agents cannot change group, so that for all $i$,
the total number $N_i=\sum_q n_{qi}$ of agents of group $i$ is fixed.
The satisfaction of agents of type $i$ on site $q$ is characterized by
a utility $U_{qi}(n_{q1},...,n_{qm})$ that depends only on the numbers
of agents of each group on the same site $q$.

The model is defined with a continuous time dynamics following a logit (or Glauber) rule,
which is commonly used in social sciences and in particular economic works \cite{Anderson92}.
If transitions between sites $q$ and $q'$ are allowed,
agents move from $q$ to $q'$ with a probability per unit time
\begin{equation} \label{logit}
W = \frac{\nu_0}{1 + e^{-\Delta U / T}},
\end{equation}
where $\Delta U = U_{q'i}'-U_{qi}$ is the variation of the agent's own utility, with
\begin{eqnarray}
U_{q'i}' &=& U_{q'i}(n_{q'1},...,n_{q'i}+1,...,n_{q'm})\\
U_{qi} &=& U_{qi}(n_{q1},...,n_{qi},...,n_{qm}).
\end{eqnarray}
The parameter $T$ plays the role of an effective
temperature, introducing noise in the decision process to take into account
other factors influencing choices \cite{Anderson92},
and $\nu_0$ is a characteristic transition frequency.

In order to obtain analytical results, we assume that the utility function
is such that the change of individual utility experienced
by an agent during a move can be expressed
as the variation of a function of the global configuration
$x=\{n_{qi}\}$ \cite{pnas2009}. More precisely,
we assume that there exists a function $L(x)$ such that for each agent
in group $i$, moving from node $q$ to node $q'$, 
\begin{equation} \label{dUdL}
U_{q'i}' - U_{qi} = L(y)-L(x)
\end{equation}
where $y=(n_{11},\ldots,n_{qi}-1,\ldots,n_{q'i}+1,\ldots,n_{Qm})$ and $x=(n_{11},\ldots,n_{Qm})$
are the configurations of the system after and before the move respectively.
Such a function $L(x)$ thus provides a link between the individual behaviour
of agents and the evolution of the whole system.
In physical terms, it can be thought of as an effective energy.
The relevance of this assumption (which bears some similarities with potential games \cite{Monderer})
for the general class of systems considered above will be discussed at the end of the paper.

The stationary probability distribution
$\mathcal{P}_\mathrm{s}(\{ n_{qi} \})=\mathcal{P}_\mathrm{s}(x)$
is obtained by solving the master equation governing the dynamics of the system
\cite{VanKampen}.
If Eq.~\eqref{dUdL} holds, detailed balance is satisfied \cite{VanKampen,Evans}, and we obtain
the following expression for the distribution $\mathcal{P}_\mathrm{s}(x)$:
\begin{equation} \label{dist-pn}
\mathcal{P}_\mathrm{s}(x) = \frac{1}{Z_\mathrm{s}}\, \frac{e^{L(x)/T}}{\prod_{q,i} n_{qi} !} \,
\prod_i \delta\left( \sum_q n_{qi} - N_i\right)
\end{equation}
where $Z_\mathrm{s}$ is a normalization constant.
The product of Kronecker $\delta$ functions
accounts for the conservation of the total number of agents in each group.
The different factors appearing in Eq.~(\ref{dist-pn}) can be given a simple interpretation.
The exponential factor directly comes from the detailed balance associated to the logit
rule Eq.~(\ref{logit}), while the product of factorials appearing at the denominator in
Eq.~\eqref{dist-pn} results from the coarse-graining of configurations.
Namely, given the numbers of agents $\{n_{qi}\}$, there are for each group
$N_{i}!/\prod_q n_{qi}!$ ways to arrange the agents of the group.
As the numbers $N_i$ are fixed, $N_i !$ can be reabsorbed
into the normalization constant.

Defining a density of agents $\rho_{qi}=n_{qi}/H$,
where $H \gg 1$ is a characteristic number (for instance a maximal number of agents on a site),
the utility $U_{qi}$ then becomes a function $u_{qi}(\rho_{q1},...,\rho_{qm})$.
We further assume that the function $L(x)$ can be written in the large deviation
form \cite{Touchette}
\be \label{Ltilde}
L(x) = H \tilde{L}(\{\rho_{qi}\}).
\ee
To determine $\tilde{L}$, we combine Eqs.~\eqref{dUdL} and \eqref{Ltilde},
and expand $\tilde{L}$ to leading order in $1/H$, yielding
\be
\frac{\partial \tilde{L}}{\partial \rho_{q'i}} -
\frac{\partial \tilde{L}}{\partial \rho_{qi}}
= u_{q'i} - u_{qi}.
\ee
By identification, we get for all $q$
\be \label{eq-dLdrho}
\frac{\partial \tilde{L}}{\partial \rho_{qi}} =
u_{qi}(\rho_{q1},...,\rho_{qm}).
\ee
As the r.h.s.~of Eq.~(\ref{eq-dLdrho}) only depends on densities of agents on node $q$,
$\tilde{L}$ necessarily takes the form
\be
\tilde{L}(\{\rho_{qi}\}) = \sum_q l_q(\rho_{q1},...,\rho_{qm}),
\ee
and one has
\be \label{lq-uq}
\frac{\partial l_{q}}{\partial \rho_{qi}} = u_{qi}.
\ee
If there is a single group ($m=1$), $l_{q}(\rho_q)$ is simply obtained by
integrating $u_q(\rho_q)$. In contrast, if $m>1$, $l_{q}$ (and thus $\tilde{L}$) only exists
if the following condition, resulting from the equality of cross-derivatives
of $l_{q}$, is satisfied:
\be \label{cross-deriv}
\frac{\partial u_{qi}}{\partial \rho_{qj}} =
\frac{\partial u_{qj}}{\partial \rho_{qi}}, \qquad i\ne j.
\ee
If this condition holds, the stationary distribution reads, after an expansion of the
factorials using Stirling's formula,
\be \label{dist-can}
\mathcal{P}(\{\rho_{qi}\}) = \frac{1}{Z} \prod_q e^{H f_q(\rho_{q1},...,\rho_{qm})/T} \prod_i \delta \Big(\sum_{q} \rho_{qi} - Q \overline{\rho}_i \Big)
\ee
where $f_q$ is given by
\be \label{eq-fq}
f_q(\rho_{q1},...,\rho_{qm}) = l_q(\rho_{q1},...,\rho_{qm})
+T s(\rho_{q1},...,\rho_{qm}),
\ee
with
\be
\label{expr-entropie}
s(\rho_{q1},...,\rho_{qm}) = - \sum_i \rho_{qi} \ln \rho_{qi}.
\ee
In analogy to physical systems, $f_q(\rho_{q1},...,\rho_{qm})$
can be interpreted as a local free
energy (up to a change of sign), and the term $s(\rho_{q1},...,\rho_{qm})$,
which is multiplied
by the 'temperature' $T$, may be seen as an entropic contribution
associated to the node $q$.

%%%%%%%%%%%%%%%%%%%%%%%%%%%%%%%%%%%%%%%%%%%%%%%%%%%%%%%%%%%%%%%%%%%%%%%
\section{Utility and chemical potential}

We now turn to the main result of this letter.
The configurations $\{ \rho_{qi}^* \}$ which maximize $F=\sum_q f_q$ under the constraints of fixed global density $\sum_q \rho_{qi} = Q \overline{\rho}_i$ are the most probable (or equilibrium)
configurations. Finding the equilibrium densities of agents is then a constrained maximization problem. Let us introduce a Lagrangian
\bea \label{lagrange}
\mathcal{L}(\{ \rho_{qi} \},\{\lambda_i\}) &=& \sum_q f_q(\rho_{q1},...,\rho_{qm})\\ \nonumber
&& \qquad \qquad - \sum_i \lambda_i \Big(\sum_{q} \rho_{qi} - Q \overline{\rho}_i \Big),
\eea
where the parameters $\lambda_i$ are Lagrange multipliers associated to
the conservation of the number of agents in each group.
In physical terms, $\lambda_i$ corresponds to the chemical
potential\footnote{An equivalent formulation is to define the chemical
potential $\lambda_i$ as the logarithmic derivative of the partition function
$Z$ with respect to $N_i$, a definition that can be extended to some classes of
nonequilibrium models \cite{Bertin07}.
Note also that the standard definition of chemical potential for equilibrium systems
differs by a conventional factor $-1/T$ from the one we use here
\cite{balescu}.} of the agents of group $i$.
The equilibrium densities $\{ \rho_{qi}^* \}$ are then determined from the conditions
$\partial \mathcal{L} / \partial \rho_{qi}=0$ for all $(q,i)$, yielding
\begin{equation} \label{max-lagrange}
u_{qi}(\rho_{q1}^*,...,\rho_{qm}^*) + T \frac{\partial s}{\partial \rho_{qi}}(\rho_{q1}^*,...,\rho_{qm}^*) = \lambda_i,
\end{equation}
which is the main result of this letter.
Equation (\ref{max-lagrange}) thus provides an answer to the question raised at the beginning of
this letter: there is indeed a direct relationship between the socio-economic utility and the chemical potential defined, in analogy to equilibrium physical systems, from the conservation of the number of particles. At zero temperature, both quantities can be identified.
This result might come as a surprise: utility is often thought to be the socio-economic concept most similar to the physical concept of energy (or more precisely, the opposite of the energy), because agents seek to maximize their utility in social systems and physical particles minimize the energy in the zero temperature limit. Hence one might have intuitively expected the homogeneity of utility to be linked to a notion of temperature (the thermodynamic variable conjugated to energy), rather than to a chemical potential.

Interestingly, Eq.~(\ref{max-lagrange}) not only provides a link between two apparently
unrelated concepts, but also yields a non-trivial prediction on the variations
of utility across the system at non-zero temperature.
As the chemical potential remains uniform at any temperature, one sees from Eq.~(\ref{max-lagrange})
that the utility $u_{qi}=\lambda_i - T\partial s/\partial\rho_{qi}$
becomes non-uniform if $T>0$, and that the corrections to
uniformity are given by the derivative of the local entropy.

In a statistical physics language, Eq.~\eqref{dist-can} corresponds to the
canonical ensemble, where the number of interacting entities (agents or
particles) is fixed. It is sometimes convenient to consider the
so-called grand-canonical ensemble, where particles are exchanged with
an external reservoir. In the context of agent-based models,
the reservoir corresponds to the external world.
This means that we implicitly consider a very large set of sites ('the world')
and focus only on a small subpart of it ('the system'), still containing
a large number of agents.
Since the 'world' has a fixed number of agents, it can be described by the
stationary distribution Eq.~(\ref{dist-pn}).
Following standard statistical physics methods \cite{balescu}, the probability distribution
of the considered subpart is given by
\be \label{dist-gdcan}
\mathcal{P}_\mathrm{ow}(\{\rho_{qi}\}) = \frac{1}{Z_\mathrm{ow}}
\prod_q e^{H[f_q(\rho_{q1},...,\rho_{qm}) -\sum_i \lambda_i \rho_{qi}]/T},
\ee
where $\lambda_i$ is the chemical potential of group $i$
imposed by the external world.
Finding the most probable densities $\rho_{qi}^*$ is now straightforward since the
densities on different sites are independent.
Maximizing the argument of the exponential in Eq.~(\ref{dist-gdcan}),
one recovers Eq.~(\ref{max-lagrange}).

In the following, we give two examples of models belonging to the above generic class,
in the context of urban economics.

%%%%%%%%%%%%%%%%%%%%%%%%%%%%%%%%%%%%%%%%%%%%%%%%%%%

\section{A simple urban economics model}
The model presented here is a simple model of land use and transport interaction
in urban economics \cite{Fujita89}.
In this model, a city is described as a grid composed of $Q$ blocks. 
In each block, one or several agents (representing households) can live by paying a
rent to the landowner. 
A central business district (CBD) is placed on the grid and all agents commute there for their work (monocentric city model). A transport cost $c$ per unit distance is associated to this commuting. The distance between a block $q$ and the CBD is denoted by $r_q$. 
The size of the city is fixed: a constant radius $r_f$ defines the urban fringe, out of which no agent lives\footnote{In standard urban economics models, land is used for agriculture
outside the city, and the landowners then earn an agricultural rent
\cite{Fujita89}. These landowners rent to the highest bidder,
so that all prices must be greater than the agricultural rent.
However, to simplify the presentation, we have dropped the agricultural
rent parameter by introducing a fixed city size.}.
All agents have the same income $Y$, which is spent on transport, on housing
and on a composite good $z$ representing all other consumer goods.
This gives a budget constraint for each agent
\begin{equation} \label{budget}
Y = z + cr_q + \sigma p_q
\end{equation}
where $\sigma$ is the surface of housing, and $p_q$ is the rent per unit surface in block $q$.
We first consider a simple model where all agents have the same surface of housing.
Each block of the grid is composed of $H$ cells of surface $\sigma_0$. 
A configuration of the city is then given by the number of agents $n_q$ in each block $q$.
We make the further simplifying assumption that the price $p_q$ of housing
in a block $q$ only depends on the density $\rho_q=n_q/H$ of agents in this block,
namely $p_q=p(\rho_q)$.
Let us emphasize that this hypothesis is an important simplification
with respect to standard urban economics models, in which the price emerges
directly from the competition for land between agents, and the density
from their utility maximization with respect to the surface of housing \cite{Fujita89}.
In cases where an explicit expression is required, we will use a logarithmic form 
\be \label{eq-prho}
p(\rho_q)= p_0 \ln (1+\rho_q),
\ee
where $p_0$ is a positive constant. 

The utility function has to be specified explicitly.
It should be an increasing function $U(z)$ of the quantity of composite good $z$ each agent
consumes, that we choose to be simply $U(z)=z$.
This means that, in the limit $T\to 0$, each agent wants to maximize the share of his income which is left after transport and housing expenses.
Using Eqs.~(\ref{budget}) and (\ref{eq-prho}), the utility $U$ becomes a function
$u_q(\rho_q)$ of the local density,
\begin{equation}
u_q(\rho_q) =  Y-cr_q-\sigma_0 p(\rho_q).
\end{equation}

Urban economics distinguishes closed city models, where the total number $N$ of agents is fixed,
and open city models, where $N$ fluctuates
due to exchanges with the external world \cite{Fujita89}. 
We start by considering the closed city model.
In the continuous limit where $H$ and $N\rightarrow \infty$ with the average density
$\overline{\rho} = N/(HQ)$ fixed, the stationary probability distribution
takes the form Eq.~\eqref{dist-can}, with $f_q(\rho_q)$
given by
\be
f_q(\rho_q) = \int_0^{\rho_q} u_q(\rho)d\rho + T s(\rho_q) \\
\ee
and $s(\rho_q)=-\rho_q\ln\rho_q$.

The most probable density $\rho_q^*$ is then obtained as a function
of $\lambda$ from Eq.~\eqref{max-lagrange}, namely
\be \label{max-cc}
u_q(\rho_q^*) + T \frac{ds}{d\rho_q}(\rho_q^*) = \lambda.
\ee
In the limit $T \to 0$, often considered in socio-economic models, one finds
$\rho_q^*=\rho^*(r_q,\lambda)$, with
\begin{equation} \label{prof-rho}
\rho^*(r_q,\lambda) = p^{-1}\left( \frac{Y-cr_q-\lambda}{\sigma_0} \right),
\end{equation}
where $p^{-1}$ is the reciprocal function of $p$.

The parameter $\lambda$ is then determined from the density constraint
$\sum_q \rho_q^* = Q \overline{\rho}$.
Following standard literature \cite{Fujita89},
we focus here on the simplest situation of a one-dimensional city.
Using the continuous approximation
\begin{equation}
\frac{1}{Q}\sum_q \rho_q^* \approx \frac{1}{r_f} \int_{0}^{r_f}\rho^*(r,\lambda)\, dr,
\label{integration}
\end{equation}
we compute the average density $\overline{\rho}(\lambda)$, and then determine numerically
the reciprocal function $\lambda(\overline{\rho})$.

We now briefly turn to the open city model (similar to the above 'open world' case)
where agents can also move to or from a large number of other cities.
The stationary distribution is given by Eq.~\eqref{dist-gdcan},
which in the present open city model simplifies to
\be \label{dist-opencity}
\mathcal{P}_\mathrm{oc}(\{\rho_{qi}\}) = \frac{1}{Z_\mathrm{oc}}
\prod_q e^{H[f_q(\rho_q) - \lambda \rho_q]/T}.
\ee
Finding the most probable density is then an unconstrained maximization
problem. The relation $df_q/d \rho_q = \lambda$ yields the same
equation as \eqref{max-cc}, resulting in the same density profile
\eqref{prof-rho} in the limit $T \to 0$.
For $T>0$, the density can be obtained from a numerical resolution of Eq.~(\ref{max-cc}).
We find that increasing the temperature $T$ progressively blurs the zero temperature
profile given by Eq.~(\ref{prof-rho}), eventually leading to a homogeneous density.
The same effect has been observed in urban economics models \cite{Anas90}.
As a consequence, the city is more spread, leading to a utility gain for agents near the city
center, and to a loss for agents in the periphery.

%%%%%%%%%%%%%%%%%%%%%%%%%%%%%%%%%%%%%%%%%%%%%%%%%%%
%%%%%%%%%%%%%%%%%%%%%%%%%%%%%%%%%%%%%%%%%%%%%%%%%%%

\section{Urban model with two types of agents}

In this second model, two income groups are distinguished.
Rich agents (group $1$) have an income $Y_1$ and a surface of housing $\sigma_1$,
while poor agents (group $2$) have an income $Y_2<Y_1$ and a surface of housing
$\sigma_2<\sigma_1$. Each block contains at most $H$ agents, irrespective of their group.
A configuration of the city is described by the densities $\rho_{q1}=n_{q1}/H$ and $\rho_{q2}=n_{q2}/H$ in each block $q$.
The price $P_{qi}$ an agent of group $i$ pays for housing in block $q$ depends on his surface of housing and on the local density of poor and rich agents:
\begin{equation}
\label{prices}
\begin{array}{l}
P_{q1}(\rho_{q1},\rho_{q2}) = \sigma_1\, \tilde{p}(a_1 \rho_{q1} + b_1 \rho_{q2})\\
P_{q2}(\rho_{q1},\rho_{q2}) = \sigma_2\, \tilde{p}(a_2 \rho_{q1} + b_2 \rho_{q2})
\end{array}
\end{equation}
where $a_1$, $b_1$, $a_2$ and $b_2$ are given constants, and $\tilde{p}$ a function
to be determined.
The utility function of an agent of group $i=1,2$ in block $q$ has the form
\begin{equation}
u_{qi}(\rho_{q1},\rho_{q2}) = Y_i - c r_q - P_{qi}(\rho_{q1},\rho_{q2}).
\end{equation}

\begin{figure}[t]
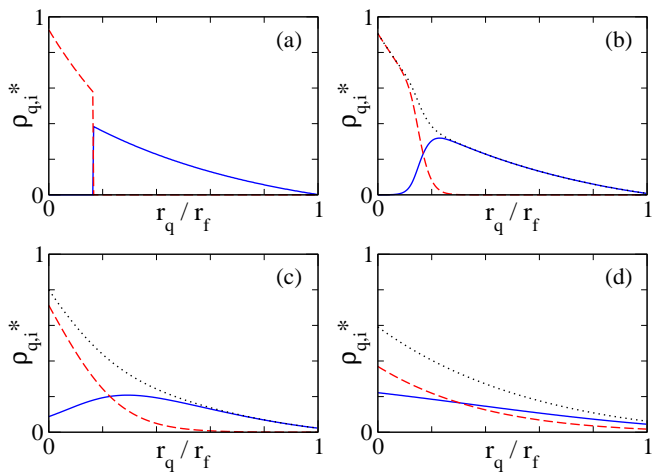

\onefigure[width=8.5cm]{eplfig1b.eps}
\caption{Density profile $\rho_{qi}^*$ as a function of $r_q$ for both groups of agents
(rich, full line; poor, dashed line) for different temperatures:
$T/T_0'=0.0018$ (a), $0.018$ (b), $0.089$ (c) and $0.36$ (d),
with $T_0' \equiv p_0 \sigma_2$.
The dotted lines indicate the total density $\rho_{q1}^* + \rho_{q2}^*$. Parameters:
$p_0=1.4$, $c=0.4$, $Y_1=452$, $Y_2=301$, $\sigma_1=6$, $\sigma_2=4$, $r_f=30$,
$\overline{\rho}_1=\overline{\rho}_2=0.13$.}
\label{Pot-2cat}
\end{figure}

The model is analytically solvable if Eq.~(\ref{cross-deriv}) is satisfied.
For this condition to hold, one can choose $a_1=a_2$ and $b_1=b_2$.
Then if $\sigma_1 b_1 = \sigma_2 a_2$, the function $\tilde{p}$ can take any form, for instance the logarithmic
form Eq.~(\ref{eq-prho}) used in the previous model, in which case we get
(choosing $b_1=\sigma_2$ and $a_2=\sigma_1$)
\begin{equation}
\label{prices2}
P_{qi}(\rho_{q1},\rho_{q2}) = \sigma_i\, p_0\ln(1+\sigma_1\rho_{q1}+\sigma_2\rho_{q2}).
\end{equation}
The stationary distribution is given by Eq.~(\ref{dist-can}), with
\be
f_q(\rho_{q1},\rho_{q2}) = l_q(\rho_{q1},\rho_{q2})
+T s(\rho_{q1},\rho_{q2}).
\ee
The expression of $s(\rho_{q1},\rho_{q2})$ is given by Eq.~(\ref{expr-entropie}), with $m=2$.
Expressing $l_q(\rho_{q1},\rho_{q2})$ explicitly, we get
\be
\label{eq-lq-2typag}
l_q(\rho_{q1},\rho_{q2})=\int_0^{\rho_{q1}} u_{q1}(\rho,0) \,d\rho
+ \int_0^{\rho_{q2}} u_{q2}(\rho_{q1},\rho) \,d\rho.
\ee
The validity of Eq.~(\ref{lq-uq}), as well as the symmetry of
Eq.~(\ref{eq-lq-2typag}) with respect to $\rho_{q1}$ and $\rho_{q2}$,
can be checked using Eq.~(\ref{cross-deriv}).
The equilibrium densities $(\rho_{q1}^*,\rho_{q2}^*)$ are determined from
Eq.~(\ref{max-lagrange}), yielding a system of two non-linear equations,
to be solved numerically.
The results of this numerical resolution are shown on Fig.~\ref{Pot-2cat}.
One recovers at low temperature the standard separation, typical of north-american cities,
between poor agents in the city center, and rich agents in the periphery \cite{Fujita89}.
The effect of a temperature increase is mainly to blur the zero temperature
pattern, hence avoiding total segregation.

Therefore, Eq.~(\ref{max-lagrange}) provides a direct prediction
for the utility profile at arbitrary temperature $T$.
It would be interesting to know whether this result remains valid beyond its a priori domain
of validity, namely for models satisfying Eq.~(\ref{cross-deriv}) so that
a function $\tilde{L}$ can be defined.
Considering again the above urban model with two types of agents,
we keep the logarithmic form Eq.~(\ref{eq-prho}) for $\tilde{p}$,
and choose as an example $a_1=a_2=1$ and $b_1=b_2=0$.
These values imply $\sigma_1 b_1 \ne \sigma_2 a_2$
so that Eq.~(\ref{cross-deriv}) is not satisfied, ruling out the possibility
to find a potential function $\tilde{L}$ and to get a simple analytical solution
of the model.

\begin{figure}[t]
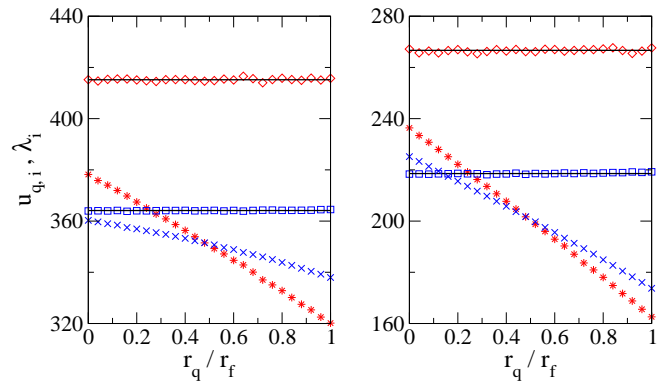

\onefigure[width=8.5cm]{eplfig2b.eps}
\caption{Agent-based simulations of the model with two types of agents
(left: $i=1$; right: $i=2$),
when no potential function $\tilde{L}$ exists (one-dimensional closed city).
The utility $u_{qi}$ is plotted as a function of $r_q$
for two temperatures ($T=100$, $*$; $T=20$, $\times$). 
Adding to the utility the term
$T\partial s/\partial \rho_{qi}$ yields the chemical potential $\lambda_i$
($T=100$, $\Diamond$; $T=20$, $\square$),
found to be constant throughout the system although the analytical solution
is not known. Full lines are horizontal lines indicating the spatially averaged value
of the chemical potential.
Parameters: $p_0=70$, $c=2$, $Y_1=480$, $Y_2=300$, $\sigma_1=6.4$, $\sigma_2=4$,
$r_f=50$, $N_1=N_2=2000$, $H=200$, $Q=51$.}
\label{potchim}
\end{figure}

Performing numerical simulations of this agent-based model
with two income groups, in the case of a one-dimensional closed city,
we first validate the simulation thanks to a comparison with the above solvable case.
Turning to the non-solvable case, we test the validity of Eq.~(\ref{max-lagrange}),
that is, whether the chemical potentials
$\lambda_i = u_{qi} + T \partial s / \partial \rho_{qi}$ ($i=1,2$)
are uniform over the city for $T>0$ (when $T\to 0$, the utility should be uniform anyhow).
We indeed observe that for a non-zero temperature, the chemical potentials are
homogeneous even in this non-solvable model, while the utility is not
(see Fig.~\ref{potchim}). We also note that although the number of agents is relatively small
($H=200$), no systematic space-dependent correction to the chemical potential is observed.

\begin{figure}[t]
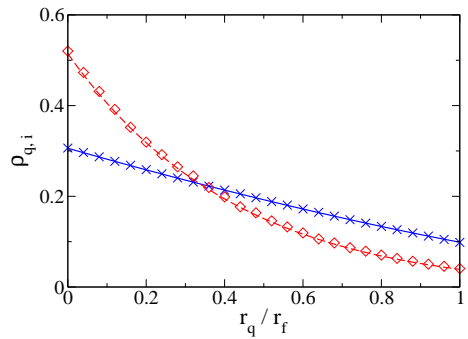

\onefigure[width=6cm]{eplfig3b.eps}
\caption{Comparison of the density profiles obtained by agent-based simulations
($i=1$, $\times$; $i=2$, $\diamond$)
and by numerical integration ($i=1$, full line; $i=2$, dashed line)
using the values of $\lambda_1$ and $\lambda_2$
measured in the agent-based model (see text).
Same parameters as in Fig.~\ref{potchim} with $T=20$.}
\label{rho-agent}
\end{figure}

We further use the obtained values of $\lambda_1$ and $\lambda_2$ to perform
a numerical resolution of the system of non-linear equations (\ref{max-lagrange}),
which we assume to be valid even in the absence of a potential $L$ function.
Interestingly, the results of the agent-based model and of the numerical resolution
are found to be in very good agreement (see Fig.~\ref{rho-agent}),
thus yielding a complementary test of the validity of the chemical potential approach.

The validity of Eq.~(\ref{max-lagrange}) in this case
can be understood as follows. In this paper, we focused on cases when
the probability distribution has the factorized form Eq.~(\ref{dist-can}),
which is a consequence of the existence of
a potential function $\tilde{L}$.
When no function $\tilde{L}$ exists, the stationary distribution is no longer
factorized, and we do not know its functional form.
However, if the stationary distribution has only short range correlations,
a chemical potential can still be introduced, in the same way as a chemical potential
can be defined in a physical system with short-range interactions \cite{balescu}.

This result is also consistent with a phenomenon known in the nonequilibrium statistical
physics literature as ``restoration of detailed balance'' \cite{Tauber,Bertin04}.
Namely, starting from a microscopic stochastic dynamics which does not obey detailed balance,
a coarse-graining procedure can lead to a detailed balance relation in terms of the effective,
coarse-grained, degrees of freedom. 
This phenomenon can appear in physical systems in cases where no macroscopic flux is present
(like fluxes of energy or particles between boundary reservoirs), as macroscopic fluxes
cannot be smeared out by the coarse-graining procedure.
In the socio-economic models we consider in this paper, there is no obvious macroscopic flux,
which suggest that detailed balance may be restored on large scales.

%%%%%%%%%%%%%%%%%%%%%%%%%%%%%%%%%%%%%%%%%%%%%%%%%%%

\section{Discussion}

In this letter, we have provided a clear relationship between the apparently
unrelated notions of socio-economic utility and chemical potential.
More specifically, we have shown that the uniformity of utility across
the social system can be traced back
to the conservation of the number of agents.
This result not only provides a conceptually interesting link, but also
yields non-trivial and testable predictions on the variations of utility among
choices (e.g., sites, blocks) when $T>0$.
We also found numerical evidence that our result
extends beyond the class of models in which it was initially derived. It would thus be
interesting to explore further its validity through numerical simulations of more
realistic models.

The idea of a non-uniform utility at equilibrium (Fig.~\ref{potchim}) may
be counter-intuitive for economists. Indeed, Nash equilibrium for
homogeneous agents implies that all have the same utility, which seems not to be the
case here when $T>0$, since agents in the border of the city have a lower utility than
those at the center. However, when noise is introduced in the decision process, a static
equilibrium picture is no longer valid. Noise allows agents to
explore the city, so that the {\it time average} value of utility is the same
for all agents, leading to a macrostate
described by Eq.~(\ref{max-lagrange}) through the ergodic hypothesis
linking time and ensemble averages.
The average utility of agents is then a decreasing function of $T$.
Note that this picture of a ``time-averaged agent'' is close in spirit to the notion
of ``representative agent'' advocated in discrete choice theory \cite{Anderson92}.
It would be interesting to investigate further the relation between these two approaches.

Another interpretation of our result is to consider the chemical potential $\lambda_i$ as an effective utility. We first note that the distribution $\mathcal{P}(\{\rho_{qi}\})$ at $T>0$
can be obtained from the zero-temperature distribution by replacing $l_q$ by
$f_q=l_q+Ts$ [see Eqs.~(\ref{dist-can}) and (\ref{eq-fq})],
in the same way as the macroscopic energy is replaced by the free
energy in a physical system at finite temperature.
Then, changing $l_q$ into $f_q$ in Eq.~(\ref{lq-uq}), we get an effective
utility $u_{qi}^{\mathrm{eff}}=\partial f_q/\partial\rho_{qi}$.
Hence the Nash equilibrium of an assembly of fictitious agents having this utility
would precisely correspond to Eq.~(\ref{max-lagrange}),
namely $u_{qi}^{\mathrm{eff}} = \lambda_i$.

%%%%%%%%%%%%%%%%%%%%%%%%%%%%%%%%%%%%%%%%%%%%%%%%%%%

\acknowledgments
Fruitful discussions with S.~Grauwin and F.~Goffette-Nagot are hereby gratefully acknowledged.

\end{document}